\documentclass[aps,pra,reprint,superscriptaddress]{revtex4-2}

\usepackage{graphicx}
\usepackage{amsmath}
\usepackage{comment}
\usepackage{braket}
\usepackage{xcolor}
\usepackage{hyperref}
\hypersetup{
    colorlinks=true,
    citecolor=teal,
    linkcolor=teal,
    filecolor=teal,      
    urlcolor=teal,
}

\usepackage[nameinlink,capitalise]{cleveref}
\newcommand{\expect}[1]{\langle #1 \rangle}

\usepackage[normalem]{ulem}

\begin{document}

\title{Structural complexity of snapshots of 2D Fermi-Hubbard systems}

\author{Eduardo Ibarra-Garc\'ia-Padilla}
\email[]{edibarra@ucdavis.edu}
\affiliation{Department of Physics and Astronomy, University of California, Davis, California 95616, USA}
\affiliation{Department of Physics and Astronomy, San Jos\'e State University, San Jos\'e, California 95192, USA}
\author{Stephanie Striegel}
\affiliation{Department of Physics and Astronomy, San Jos\'e State University, San Jos\'e, California 95192, USA}
\author{Richard T. Scalettar}
\affiliation{Department of Physics, University of California, Davis, California 95616, USA}
\author{Ehsan Khatami}
\email[]{ehsan.khatami@sjsu.edu}
\affiliation{Department of Physics and Astronomy, San Jos\'e State University, San Jos\'e, California 95192, USA}

\date{\today}

\begin{abstract}
The development of quantum gas microscopy for two-dimensional optical lattices has provided an unparalleled tool to study the Fermi-Hubbard model (FHM) with ultracold atoms. Spin-resolved projective measurements, or snapshots, have played a significant role in quantifying correlation functions, theory verification, and thus, the uncovering of underlying physical phenomena such as antiferromagnetism at commensurate filling on bipartite lattices, and other charge and spin correlations, as well as dynamical properties at various densities. Here, we employ a recent concept, the {\em multi-scale structural complexity}, and show that when computed for the snapshots (of either single spin species, local moments, or total density) it can provide a theory-free property, immediately accessible to experiments. Specifically, after benchmarking results for Ising and XY models, we study the structural complexity of snapshots of the repulsive FHM in the two-dimensional square lattice as a function of doping and temperature. We generate projective measurements using determinant quantum Monte Carlo and compare their complexities against those from the experiment. We demonstrate that these complexities are linked to relevant physical observables such as the entropy and double occupancy. Their behaviors capture the development of correlations and relevant length scales in the system. We provide an open-source code in Python which can be implemented into data analysis routines in experimental settings for the square lattice~\cite{Ibarra2023_github}.
\end{abstract}

\maketitle

\section{Introduction}

The Fermi-Hubbard model (FHM) is of great importance in describing the electronic properties of strongly-correlated materials. It accurately captures some their key characteristics, and it exhibits a number of canonical phases of matter, hosting a metal-to-insulator crossover and magnetic and charge order in the two-dimensional (2D) square lattice, and is extensively researched in relationship to $d-$wave superconductivity~\cite{white1989numerical,scalapino1999superconductivity,maier2006structure,Arovas2021,Qin2020,Qin2021}.

One of the major accomplishments in quantum simulation has been the precise engineering of the FHM with ultracold atoms in optical lattices. These exhibit flexible tunability of the kinetic and interaction energies, as well as the lattice filling or even the lattice geometry~\cite{jordens2010quantitative,paiva2010fermions,schneider2012fermionic,greif2013short,hart2015observation,boll2016spin,mazurenko2017cold,Xu2022,Jirayu2022,Lebrat2023,prichard2023}. 
In lower dimensional lattices, i.e. one-dimensional (1D) and 2D, quantum gas microscopy has provided direct observation of  correlations beyond nearest neighbors through real-space imaging of spin-resolved projective measurements~\cite{bakr2009quantum,cheuk2015quantum,parsons2016site,moses2017new,Altman2021,Gross2017,Bloch2012,Cheuk2016,Gross2021,Hartke2020,Hartke2023}. These results yield profound insights into the phase diagram~\cite{Bohrdt2021}, in particular regarding antiferromagnetism,  and how spin correlations are affected by the presence of holes at intermediate temperatures.

Despite these great achievements, many questions remain open about the model in two dimensions, in particular those pertaining to the pseudogap, strange metal, and possible superconducting regions, where complex patterns and competing orders emerge that are hard to describe quantitatively~\cite{Huang2018,Mai2023,Huang2023,xiao2023temperature}. 

One possible route to extract physical information from projective measurements in an unbiased way, where such quantitative descriptions are difficult, is by enlisting the help of machine learning or other data-driven techniques~\cite{Johnston2022}. For example, an early application of supervised learning aimed to detect the onset of the pseudogap phase upon doping in experimental snapshots~\cite{Bohrdt2019}. In another study, kernels of a convolutional neural network, trained to distinguish experimental snapshots taken at high and low temperatures, were used to analyze correlations that develop in the system at low temperatures, including those at dopings relevant to the strange metal phase~\cite{Khatami2020,Striegel2023}. Machine learning has also been successfully exploited to obtain the relation between the ground state energy and the density~\cite{nelson2019machine}.

In this work, we focus on a recently introduced unbiased measure, called the multi-scale structural complexity~\cite{Bagrov2020}, which is based on dissimilarity of patterns at different scales and uses ideas from the renormalization group flow to aggregate information about different scale correlations present in the system. So far, it has been used for analyzing snapshots of 1D classical and quantum systems~\cite{Sotnikov2022}, including systems out of equilibrium~\cite{Maletskii23}. It has also been shown to be capable of detecting phase transitions to and from the off-diagonal bond-density-wave ordered phase of the half-filled extended FHM in one dimension despite only employing diagonal (density) snapshots~\cite{Xiao2022}. Given the large datasets attainable in experiments with quantum gas microscopes, we demonstrate that the structural complexity can be an immediately useful tool to analyze Fermi-Hubbard snapshots.

We first apply the structural complexity technique to classical models of magnetism to build an intuition of what the dissimilarities and the complexities capture. We then study the structural complexity of the repulsive FHM in the 2D square lattice at intermediate coupling (where the on-site interaction equals the non-interacting bandwidth) as a function of hole doping $\delta$ and temperature $T$. This value of the interaction is where the antiferromagnetic (AFM) correlations are maximal at commensurate filling, and therefore, allows for a good comparison against classical models as well as sets the stage for the study of the doped system. On the theory side, we perform  determinant quantum Monte Carlo (DQMC)~\cite{Blankenbecler1981,Sorella1989} simulations, from which spin-resolved, local moments, and density snapshots are generated~\cite{Humeniuk2021}. We compare their complexities against those from experimental data and demonstrate they are directly linked to relevant physical observables such as the entropy and the double occupancy. Their behaviors illustrate the extent to which correlations arise, and capture relevant length scales in the system. 

The remainder of this paper is organized as follows: In Sec.~\ref{sec::Model_Methods} we present the models studied and the methods used. Then in Sec.~\ref{sec::Results} we present our main findings, first results for classical models, followed by those for the FHM. Finally, Sec.~\ref{sec::Conclusions} concludes our findings and presents an outlook for future studies. 

\section{Model and methods}\label{sec::Model_Methods} 

\subsection{Classical spin models}\label{subsec::Classical}

Since our main objective is to work with quantum gas microscope snapshots of Fermi-Hubbard models, which are \textit{images} with 0's and 1's as the pixel values, we first aim to gain some intuition at to what the structural complexity and the dissimilarities measure in well-known scenarios. For this reason, we perform Monte Carlo simulations of classical spin models~\cite{IbarraGarciaPadilla2016} in 2D,
\begin{equation}
    H = -J \sum_{\langle i,j\rangle} \textbf{s}_i \cdot \textbf{s}_j = -J \sum_{\langle i,j\rangle} \cos(\theta_i - \theta_j),
\end{equation}
in order to obtain and work with classical snapshots of spins.
Here, ${\bf s}_i$ is the spin vector at site $i$, and $\left<\dots\right>$ indicates a sum over nearest-neighbors.
We focus on the ferromagnetic (FM) Ising model ($J>0$, $\theta = \pm \pi/2$), the antiferromagnetic (AFM) Ising model ($J<0$, $\theta = \pm \pi/2$), and the FM XY model ($J>0$, $\theta \in [0,2\pi)$). Although the location of critical temperature is the same for the FM and AFM Ising models, $T_c/J = 2.269$, the magnetic orderings (and thus the relevant patterns in snapshots) are different, and therefore one expects the behavior of the structural complexities to be different as well.

On the other hand, the 2D FM XY model does not have an order parameter, nor does it exhibit long-range order. Rather, it has a finite-temperature Berezinskii–Kosterlitz–Thouless (BKT) transition at $T_c/J = 0.88$. In this model, vortices are the topologically stable configurations, and the transition is characterized by the binding-antibinding of pairs of vortices with opposite vorticity. Due to these  fundamental differences, exploring this model will yield relevant information on what the complexity measures capture in the presence of more intricate patterns and topological phase transitions. It is worth noting as well that the transition into a superfluid phase in the 2D attractive FHM at generic fillings is in the BKT class~\cite{Scalettar1989_attrac_U}, so an understanding of the structural complexity in the XY model could be directly relevant to future studies of the former model as well.

\subsection{The Fermi-Hubbard model and the DQMC}\label{subsec::FHM}

We are mainly interested in investigating the FHM, 
\begin{equation}\label{eq:Hubbard_N1}
H = -t \sum_{\langle i,j \rangle, \sigma} \left( c_{i \sigma}^\dagger c_{j \sigma}^{\phantom{\dagger}} 
+ \mathrm{h.c.} \right) + U \sum_{i} n_{i \uparrow} n_{i \downarrow},  - \mu \sum_{i,\sigma} n_{i \sigma},
\end{equation} 
where $c_{i \sigma}^\dagger$ ($c_{i \sigma}^{\phantom{\dagger}} $) is the creation (annihilation) operator for a fermion with spin flavor $\sigma = \uparrow,\downarrow$ on site $i = 1,2,...,N$ in a 2D square lattice, $N$ denotes the number of lattice sites, $n_{i \sigma} = c_{i \sigma}^\dagger c_{i \sigma}^{\phantom{\dagger}}$ is the number operator for flavor $\sigma$, $t$ is the nearest-neighbor hopping amplitude, and $U$ is the interaction strength. We work in the grand canonical ensemble and use the chemical potential $\mu$ to adjust the fermion density. We set $\hbar=k_B = 1$ throughout the paper, and consider the $U>0$ (repulsive) case.

In this work, we generate snapshots of projective measurements in $10 \times 10$ lattices using the method described in Ref.~\cite{Humeniuk2021,Humeniuk_git}, in which nested componentwise direct sampling of fermion pseudodensity matrices is used to generate an ensemble of pseudosnapshots, which, when reweighted, is equivalent to an ensemble of projective measurements of the occupation numbers. The procedure introduces a weight associated with each snapshot, which for all the cases we consider here is more or less uniform among snapshots, and we therefore ignore for simplicity.
We use DQMC with a Trotter step of $\Delta \tau = 0.05t$. The projective measurements correspond to the spin-resolved  densities $n_{i\sigma}$ in the lattice, from which the local moment $m^z_i = \vert n_{i\uparrow} - n_{i\downarrow} \vert$ and the total density $n_i = n_{i\uparrow} + n_{i\downarrow}$ are constructed. We apply point-group symmetries to increase the number of samples eightfold.  

\subsection{Structural complexity}\label{subsec::SC}

\begin{figure}[tbp!]
\includegraphics[width=0.75\linewidth,angle=270]{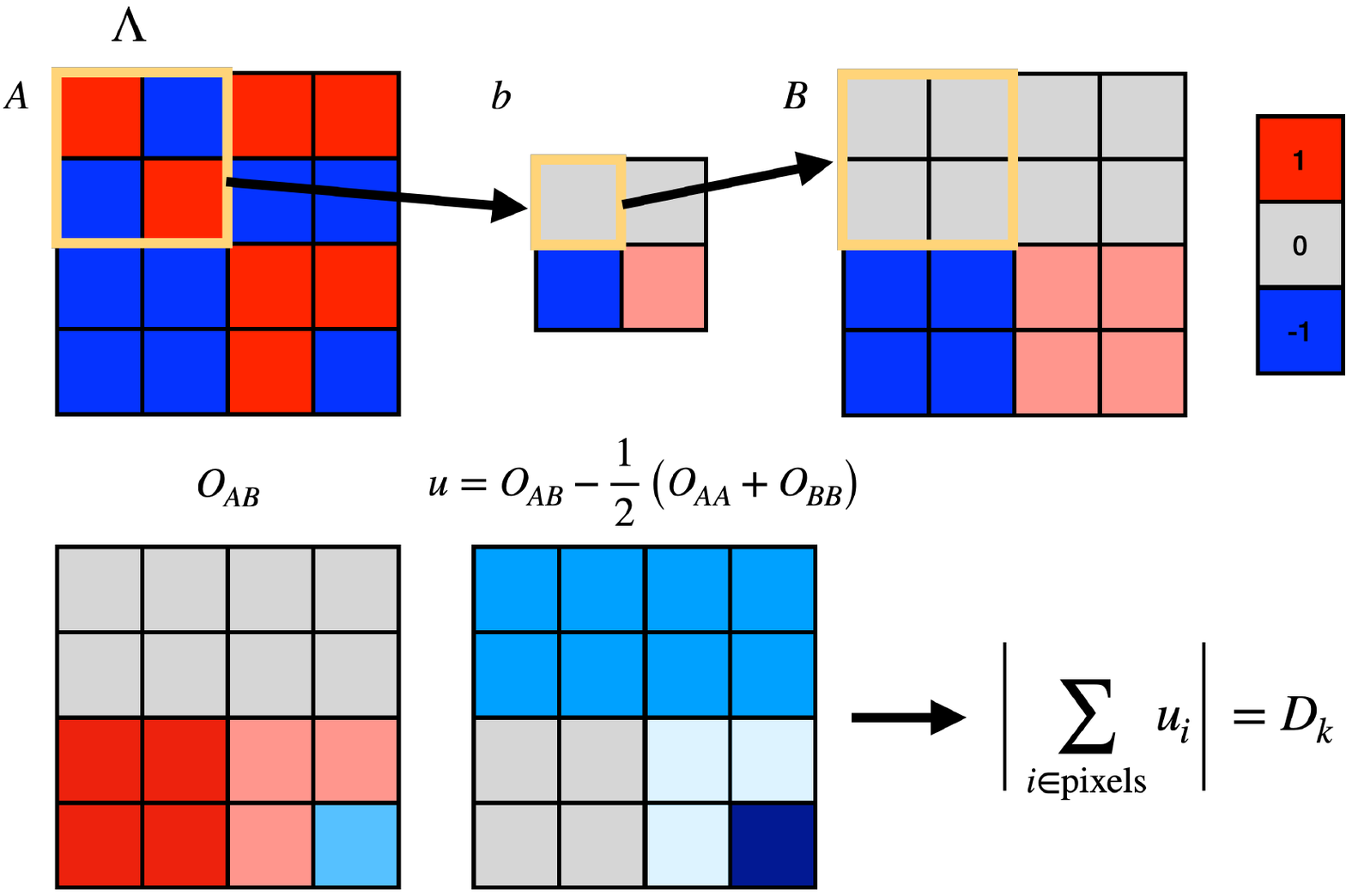}
\caption{\textbf{Structural complexity procedure.} An image $A$ is coarse-grained using a $\Lambda\times \Lambda$ window. The resulting image after coarse-graining corresponds to image $b$, which is then resized to its original size (image $B$). Overlaps between images are computed by taking the dot product of the arrays. Here, we illustrate the pixel-by-pixel product for the overlap between $A$ and $B$ ($O_{AB}$), and the pixel-by pixel result for the differences between overlaps ($O_{AB} - \frac{1}{2}[O_{AA}+ O_{BB}]$). Performing a sum over pixels and taking the absolute value of the result corresponds to the dissimilarity $D_k$.
}
\label{fig::SC_procedure} 
\end{figure}

The qualitative concept of the complexity of patterns, systems, and processes is inherent to human perception and plays an important role in natural and social sciences. A precise mathematical description was given recently~\cite{Bagrov2020} which formulates a single unique number that characterizes the \textit{structural complexity} of an image. This number is obtained via a series of coarse-graining steps (as is done in renormalization group calculations). In each step, information from different scales present in the system aggregates. Formally, the structural complexity $\mathcal{C}_0$ is defined as:
\begin{align}
    \mathcal{C}_0 &= \sum_{k=0}^{k_\mathrm{max} -1} D_k \nonumber \\
    &= \sum_{k=0}^{k_\mathrm{max}-1} \bigg\vert O_{k+1,k} - \frac{1}{2} \left( O_{k+1,k+1} + O_{k,k}\right) \bigg\vert,
    \label{eq:C0}
\end{align}
where $k_\mathrm{max}$ is the total number of coarse-graining steps, and  
\begin{equation}
    O_{k,p} = \frac{1}{N} \sum_{i=1}^N u^{(k)}_i u^{(p)}_i, 
\end{equation}
is the overlap function,
where the image has $N$ pixels,  and $u^{(k)}_i$ corresponds to the value of the pixel at site $i$ at coarse-graining step $k$. The coarse-graining procedure using a $\Lambda\times \Lambda$ window and the calculation of the overlaps is depicted in Fig.~\ref{fig::SC_procedure}. $D_k$ is the ``dissimilarity'' at coarse-graining step $k$.
It contains correlations that extend up to a linear size of $\Lambda^k$ in the original image and measures how different the images at consecutive coarse-graining steps $k$ and $k+1$ are. Here, we set $\Lambda=2$.  As we will see later, it is also useful to define $C_1 = \mathcal{C}_0 - D_0$ [i.e. we start the sum in Eq. (\ref{eq:C0}) at $k=1$ rather than $k=0$]. We will also see that the contributions fall off rather rapidly with $k$, so that in practice only a limited number of terms need be considered.

This definition of the complexity reflects the intuitive sense of what corresponds to a more \textit{complex} pattern in general, but also has been used specifically to detect finite-$T$ phase transitions in models such as the classical ferromagnetic Ising model~\cite{Bagrov2020}, detect quantum phase transitions in the 1D extended  FHM~\cite{Xiao2022}, and to distinguish quantum states in one-dimensional qubit chains~\cite{Sotnikov2022}. In contrast to numerically  expensive $n$-point correlation functions or neural network learning techniques to detect phase boundaries, the structural complexity offers a simple, unbiased, and numerically cheaper technique to do so.

To apply the technique to our DQMC snapshots, we tile a region of space using $256 \times 256$ of them, which allows us to perform the coarse-graining procedure up to 9 times ($k_{max}=9$). 
In addition, we compare our results against experimental data where possible, for which a similar tesselation procedure is performed. These correspond to the data used in Ref.~\cite{Khatami2020}.

\section{Results}\label{sec::Results}

\begin{figure*}[htbp!]
\includegraphics[width=\linewidth]{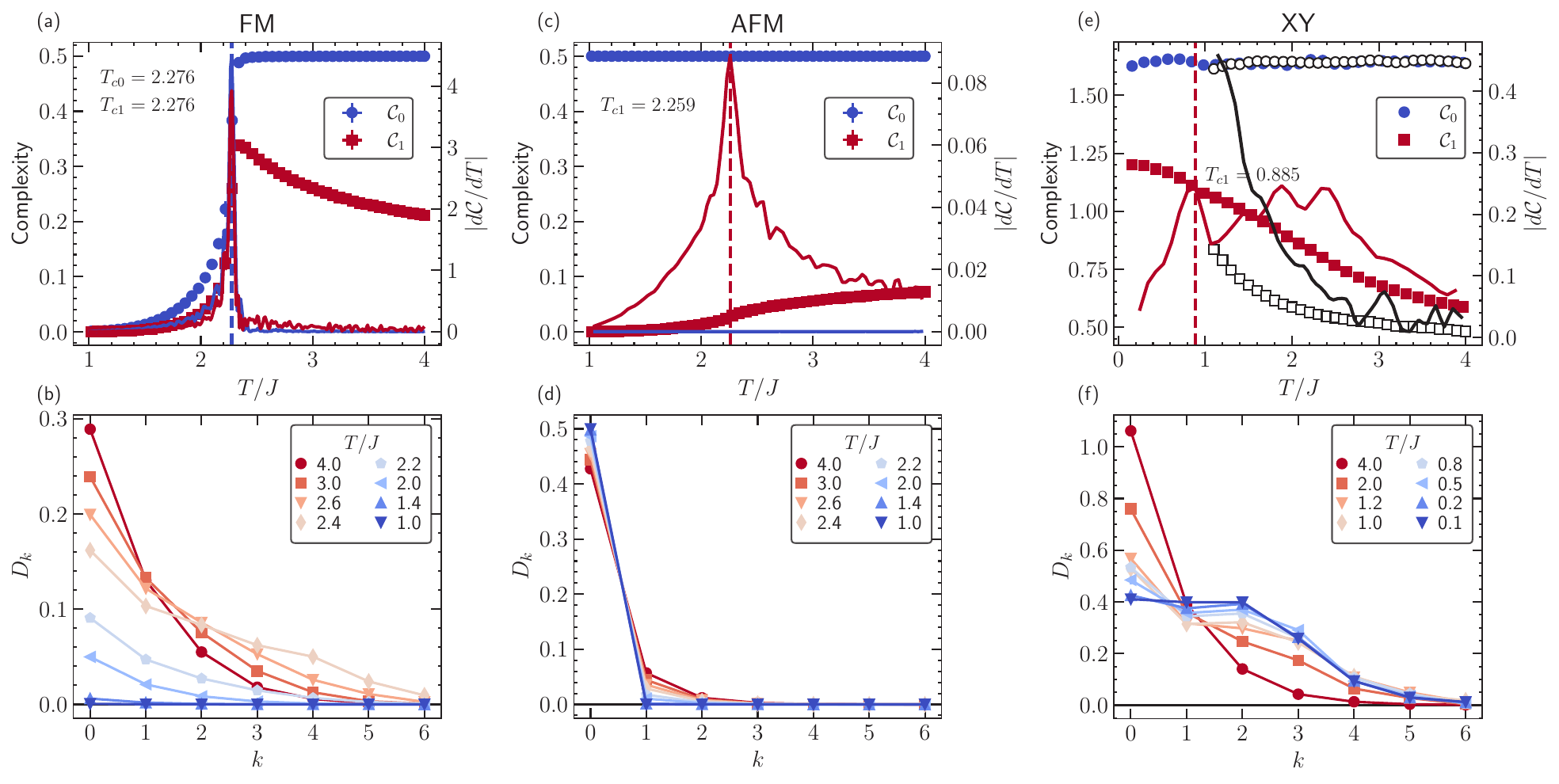}
\caption{\textbf{Classical models in $128\times128$ lattices.} Results are presented for Ising FM (a)-(b) and AFM (c)-(d) couplings, and for the XY model (e)-(f).  Upper row: $\mathcal{C}_0$ (blue circles) and $\mathcal{C}_1$ (red squares) as a function of $T/J$. Solid lines correspond to the absolute value of the derivative. The extracted transition temperatures $T_c$ are consistent with the exact results in all cases. In panel (e) black markers and lines correspond to results obtained in $256\times256$ lattices. Lower row: $D_k$ vs $k$ for different temperatures. The behavior of the dissimilarities for each model are signficantly different.}
\label{fig::Classical} 
\end{figure*}

\subsection{Classical models}\label{sub::Classical models}

\subsubsection{Hamiltonians with $\mathcal{Z}_2$ symmetry}\label{subsub::Z2}

In Fig.~\ref{fig::Classical}(a) we show the temperature dependence of the structural complexity of the FM Ising model. Our results are in agreement with Ref.~\cite{Bagrov2020} and the critical temperature, deduced from the location of the rapid drop in $C_0$ or $C_1$ or the peak in their derivatives, is in excellent agreement with the exact result. We can gain further understanding about the inner workings of the complexity measure by looking at the dissimilarities as a function of the coarse-graining step $k$ [Fig.~\ref{fig::Classical}(b)]. At high temperatures in the disordered phase, $D_k$ falls off rapidly with $k$. As the temperature is lowered to near, but still above, the transition, $D_k$ decays much more slowly, with non-vanishing values even at large values of $k$, illustrating the existence of correlations between far regions in the model. Finally, in the ordered phase, $D_k$ is mostly small and vanishes quickly at all $k$ by decreasing $T$. This makes sense since in a perfect FM, coarse-graining does not lead to a different image.

In contrast to the FM case, where  $\mathcal{C}_0$ starts at a finite but constant value and then rapidly falls off to zero below the critical temperature, we find that in the AFM Ising model, $\mathcal{C}_0$ remains constant at all temperatures, as can be seen in Fig.~\ref{fig::Classical}(c). In order to understand this, let us analyze two limiting cases. At infinite temperature, i.e.~in the fully disordered phase (where each spin orientation is equally probable), the structural complexity for both models is given by (see Appendix~\ref{App::High_T} for a derivation),
\begin{equation}\label{eq::C0_highT}
    \mathcal{C}_0 = \frac{3}{8} \left(\sum_{k=0}^\infty \frac{1}{4^k} \right) = \frac{3}{8} \left(\frac{4}{3}\right) = \frac{1}{2}.
\end{equation}  
On the contrary, in the ground state for the perfect classical FM and AFM orderings it suffices to analyze what happens after the first coarse-graining (c.g.) step, i.e. $\mathcal{C}_0 = D_0$, since it is easy to see that all other $D_k$ for $k>1$ vanish:
\begin{align}
    \mathrm{FM:} \:
    &\begin{bmatrix}
    1 & 1 \\
    1 & 1 
    \end{bmatrix}
    \xrightarrow{\text{c.g.}} 
    \begin{bmatrix}
    1 & 1 \\
    1 & 1 
    \end{bmatrix} 
    \to
    \; D_0 = 1 - \frac{1+1}{2} = 0, \\
    \mathrm{AFM:} \:
    &\begin{bmatrix}
    1 & -1 \\
    -1 & 1 
    \end{bmatrix}
    \xrightarrow{\text{c.g.}} 
    \begin{bmatrix}
    0 & 0 \\
    0 & 0 
    \end{bmatrix}
    \to
    \; D_0 = 0 - \frac{1+0}{2} = \frac{1}{2}.
\end{align}
From these two limiting cases we conclude that while in the FM case $\mathcal{C}_0$ must exhibit a drop from $1/2$ to $0$ as one lowers the temperature and the system develops a net magnetization, in the AFM case $\mathcal{C}_0$ can remain constant. We note that the variations in $C_0$ in Fig.~\ref{fig::Classical}(c) are of the order of $10^{-5}$ and not visible in the scale of the plot. Since the derivation at high-temperatures assumes an equal number of spin ups and spin downs, and in the perfect classical AFM ordering this balance is preserved at all temperatures, we infer that $\mathcal{C}_0$ is sensitive only to the uniform magnetization of the system. As we will see later, this is also the case for the spin-balanced FHM. 

On the other hand, $\mathcal{C}_1$ captures the transition temperature in the AFM Ising model with great accuracy, and exhibits a similar behavior to its FM counterpart: the complexity is larger in the disordered phase, but tends to zero in the ground state, illustrating that in the latter state, after a single coarse-graining step, the resulting patterns are no longer complex, and that there is scale invariance. 

We observe that the behaviors of the FM and AFM dissimilarities as a function of $k$ are very different. For the AFM case [Fig.~\ref{fig::Classical}(d)], in the ordered phase, only $D_0$ is nonzero and contributions at higher $k$ come from thermal fluctuations. Interestingly, because $\mathcal{C}_0$ is constant at all $T$, we observe a conservation of weights in the dissimilarities. In other words, the same exact weight loss in $D_k$ for $k>0$ is gained by $D_0$ at every temperature.

Finally, we notice that $\mathcal{D}_0$ in the FM Ising model closely tracks, within a prefactor, the temperature behavior of the entropy per spin $S$ (not shown), and $\mathcal{C}_0$ correlates with the \textit{Boltzmann entropy} $S_B$, defined as~\cite{Vijayaraghavan2017}
\begin{equation}
    S_B = -\sum_\sigma p_\sigma \log_2(p_\sigma),
\end{equation}
where $p_\sigma = (1 \pm m)/2$, and $m$ is the average magnetization of the system. That means $\mathcal{C}_1$ follows the trend in the   
\textit{total correlation density} $\rho_c= S_B - S$~\cite{Vijayaraghavan2017}.
The Boltzmann entropy assumes the lack of internal correlations and corresponds to the isolated spin entropy. Since $\rho_c$ only vanishes when no internal correlations occur, i.e. when each spin is independent, it measures how constrained the spin distribution is, and is therefore maximized when the spins are strongly correlated~\cite{Vijayaraghavan2017}.

\subsubsection{The Hamiltonian with $O(2)$ symmetry}\label{subsub::O2}

\begin{figure}[tbp!]
\includegraphics[width=\linewidth]{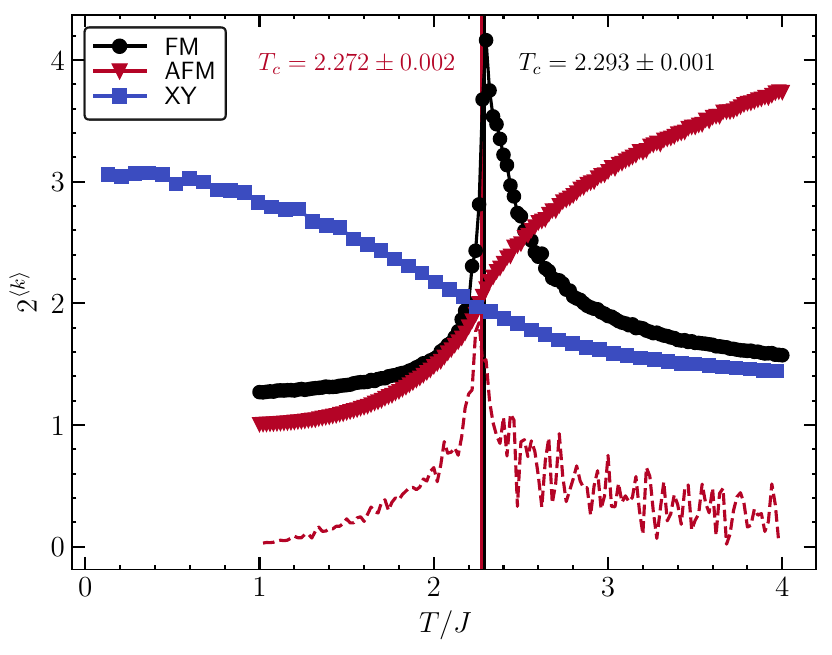}
\caption{\textbf{Average linear length scale in the classical models.} Results of $2^{\expect{k}}$ vs $T/J$ for the classical models. We apply the linear transformation $y \to 20\times(y-1)+1$ to the AFM Ising data to display their behavior on the same scale. Vertical lines indicate the location of peaks in the FM case (black) and the derivative in the AFM case (red).}\label{fig::Classical_kavg} 
\end{figure}

We now turn our attention to the XY model. In Fig.~\ref{fig::Classical}(e) we show the structural complexity of the model as a function of temperature for two system sizes $128 \times 128$ (solid markers) and $256 \times 256$ (open black markers). The results are averaged over five different initial random seeds. We further apply a moving average with a five-point window over temperature fitted with a local third-order polynomial (the Savitzky-Golay filter) to reduce noise. Similarly to the AFM Ising model, $\mathcal{C}_0$ is more or less constant for all temperatures and for both system sizes, indicating the near zero net magnetization of the system. On the contrary, $\mathcal{C}_1$ increases as the temperature is lowered, indicating that the low-$T$ phase, where the existence of pairs of vortices with opposite vorticity is expected~\cite{IbarraGarciaPadilla2016}, is the most complex one. We find a strong finite-size dependence in $\mathcal{C}_1$, expected to persist in this model for certain properties even with systems an order of magnitude larger in size~\cite{p_nguyen_21}. The derivative of $\mathcal{C}_1$ for the $128 \times 128$ system exhibits broad and inconclusive features at $T/J \lesssim 3$. By increasing the system size to $256 \times 256$, these features develop into a sharp upturn at the lowest temperatures we could access in our simulations due to inflating  autocorrelation times. This feature is consistent with the location of the peak in the specific heat ($T/J\sim 1.04$) and the BKT transition temperature of the model ($T_c/J=0.894$)~\cite{Hsieh_2013,p_nguyen_21}. In order to access lower temperatures and larger system sizes, more sophisticated numerical algorithms are required, which are beyond the scope of this work~\cite{p_nguyen_21,Prokofiev2001,Radford}.

\begin{figure*}[tbp!]
\includegraphics[width=0.9\linewidth]{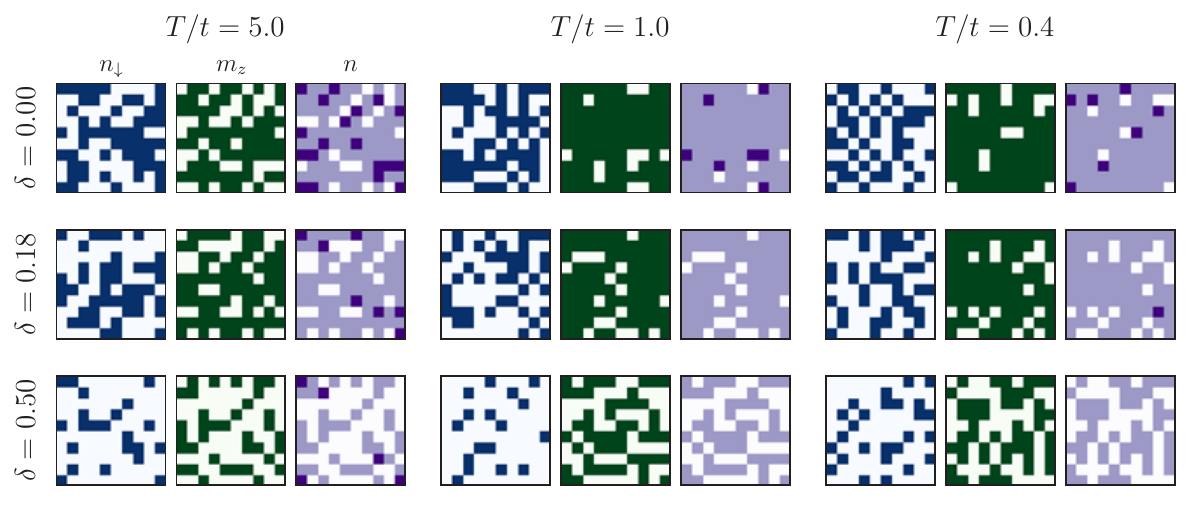}
\vspace{-0.1in}
\caption{\textbf{Sample simulated snapshots of the FHM at $U/t=8$ in $10\times10$ lattices.} Results are presented at densities $\expect{n}=1,0.82,0.5$ for temperatures $T/t = 5, 1, 0.4$ for $n_\downarrow$ densities (blue), the local moment $m_z = \vert n_\uparrow - n_\downarrow \vert $ (green) and total density $n =  n_\uparrow +  n_\downarrow $ (purple). White corresponds to 0, blue, green, and light purple to 1, and dark purple to  2. 
}\label{fig::Snapshots} 
\end{figure*}

\begin{figure}[b!]
\includegraphics[width=0.85\linewidth]{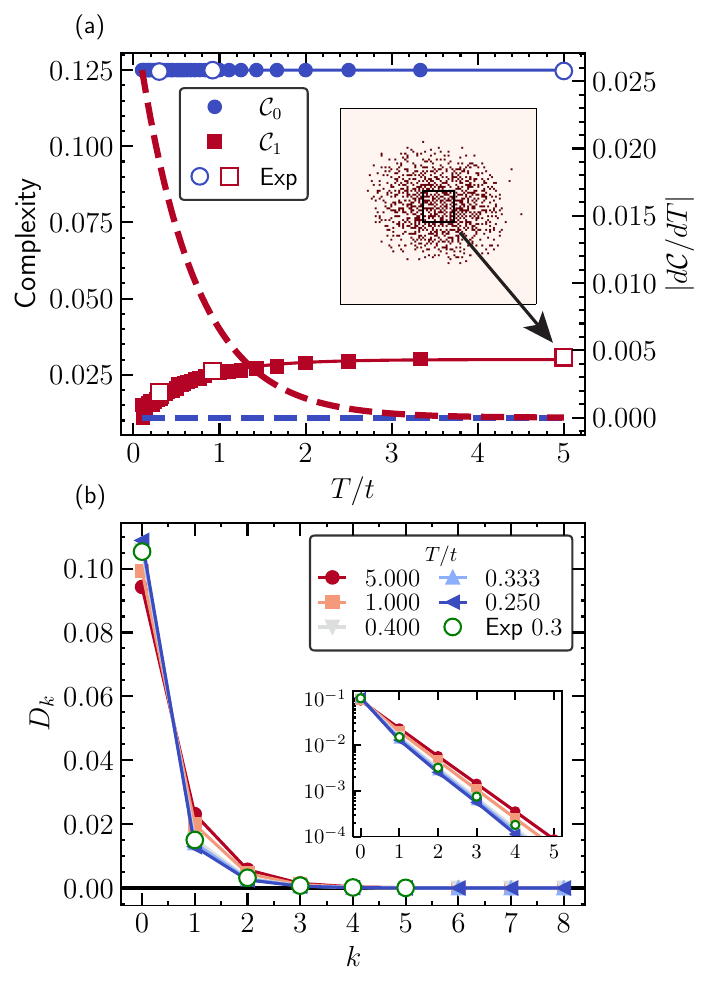}
\vspace{-0.2in}
\caption{\textbf{Structural complexity and dissimilarities of the half-filled FHM at $U/t=8$ from spin-resolved snapshots.} (a) $\mathcal{C}_0$ (blue circles) and $\mathcal{C}_1$ (red squares) from $10\times 10$ simulated snapshots as a function of temperature. Dashed lines are derivatives and open markers are based on experimental data. The inset is an example of the experimental snapshots and the $16 \times 16$ region used for the analysis. (b) Dissimilarities as a function of coarse-graining steps at different temperatures. Experimental data at $T/t=0.3$ are shown as open green circles. Inset: Same data in log scale.
}\label{fig::SC_hf_nsigma} 
\end{figure}

We find that in this case, the behavior of the dissimilarities as a function of $k$ is completely different from those for the FM or AFM Ising models as illustrated in Fig.~\ref{fig::Classical}(f). At high temperatures $D_k$ decays rapidly with $k$. However as the temperature is lowered, we find significant contributions to the complexity from all length scales for $T \lesssim T_c$, which we attribute to the existence of a non-trivial topological phase.

\subsubsection{Relevant length scales}

Since the structural complexity aggregates information about correlations in the system up to a distance of $\Lambda^{k}$ at step $k$, we define an average for $k$,
\begin{align}
    \expect{k} = \frac{\sum_k k D_k}{\sum_k D_k} = \frac{1}{\mathcal{C}_0} \sum_k k D_k,
\end{align}
and take $\Lambda^{\expect{k}}=2^{\expect{k}}$ as a relevant length scale of the model.
We therefore expect this quantity to reflect the cluster sizes needed to describe the behavior of the system at different temperatures.  

We show $2^{\expect{k}}$ vs $T/J$ in Fig.~\ref{fig::Classical_kavg} for the classical models considered here. In the case of FM Ising model, it grows rapidly near $T_c$ as the temperature is lowered and then rapidly falls in the ordered phase. Its behavior is reminiscent of the critical behavior of the magnetic susceptibility of the model. The location of the peak in this quantity (black vertical line in Fig.~\ref{fig::Classical_kavg}) provides a great estimate for $T_c$. Interestingly, we find that the behavior of $2^{\expect{k}}$ for the AFM Ising model resembles the shape of the energy, and its derivative (red dashed line in Fig.~\ref{fig::Classical_kavg}) predicts an accurate estimate for $T_c$ too (its peak location is denoted by the red vertical line). Finally, for the XY model, $2^{\expect{k}}$ grows as the temperature is lowered, indicating a growing length scale for relevant correlations, consistent with the non-trivial topological order the system exhibits. 

\subsection{The Fermi-Hubbard model}\label{sub::FHM}

In Fig.~\ref{fig::Snapshots} we present examples of the simulated snapshots of $n_\sigma,m_z$ and $n$ for $U/t=8$, three different filling fractions ($\delta = 0,0.18,0.5$) and three temperatures ($T/t = 5, 1, 0.4$). In the half-filled case, where there is on average one particle per site (first row), as the temperature is lowered, one can observe signatures of (i) Mott physics (formation of more uniform local moments), (ii) AFM (checkerboard patterns in the $n_\sigma$ snapshots), and (iii) Superexchange physics, i.e. doublon-hole quantum fluctuations which occur at neighboring sites in the $n$ snapshots, and are necessary for AFM to emerge~\cite{Cheuk2016}.
In contrast, results at 18\% doping (second row) exhibit remnants of the parent AFM state, but the presence and behavior of holes is more complicated. Finally, at 50\% doping (third row) in the Fermi liquid regime, electrons tend to avoid each other (Pauli hole) and, other than the reduction in double occupancies, there do not seem to be any significant changes as the temperature is lowered.

\subsubsection{Spin-resolved densities $n_\sigma$}\label{sec::n_sigma}

In Fig.~\ref{fig::SC_hf_nsigma}(a) we show the structural complexity of single-species snapshots of the half-filled FHM as a function of temperature. We observe that $\mathcal{C}_0$ is constant for all temperatures, similar to what was observed in the AFM Ising model, indicating a spin-balanced mixture. On the contrary, $\mathcal{C}_1$ decays monotonically as the temperature is lowered. Its derivative (dashed red line) appears to diverge as $T\to 0$ in agreement with the fact that for the 2D FHM, the N\'eel transition occurs at $T_{\mathrm{N}} = 0$. 

Results for the corresponding dissimilarities as a function of the coarse-graining step at different temperatures exhibit a similar behavior as for the AFM Ising model [see Fig.~\ref{fig::SC_hf_nsigma}(b)]. Namely, the weight lost by $D_0$ as the temperature increases is gained by $D_{k>0}$. There is however a fundamental difference here corresponding to the non-vanishing $D_1$ at all temperatures, which signals the presence of quantum fluctuations in the model. $\mathcal{C}_0$ is constant despite that.

\begin{figure*}[tbp!]
\includegraphics[width=\linewidth]{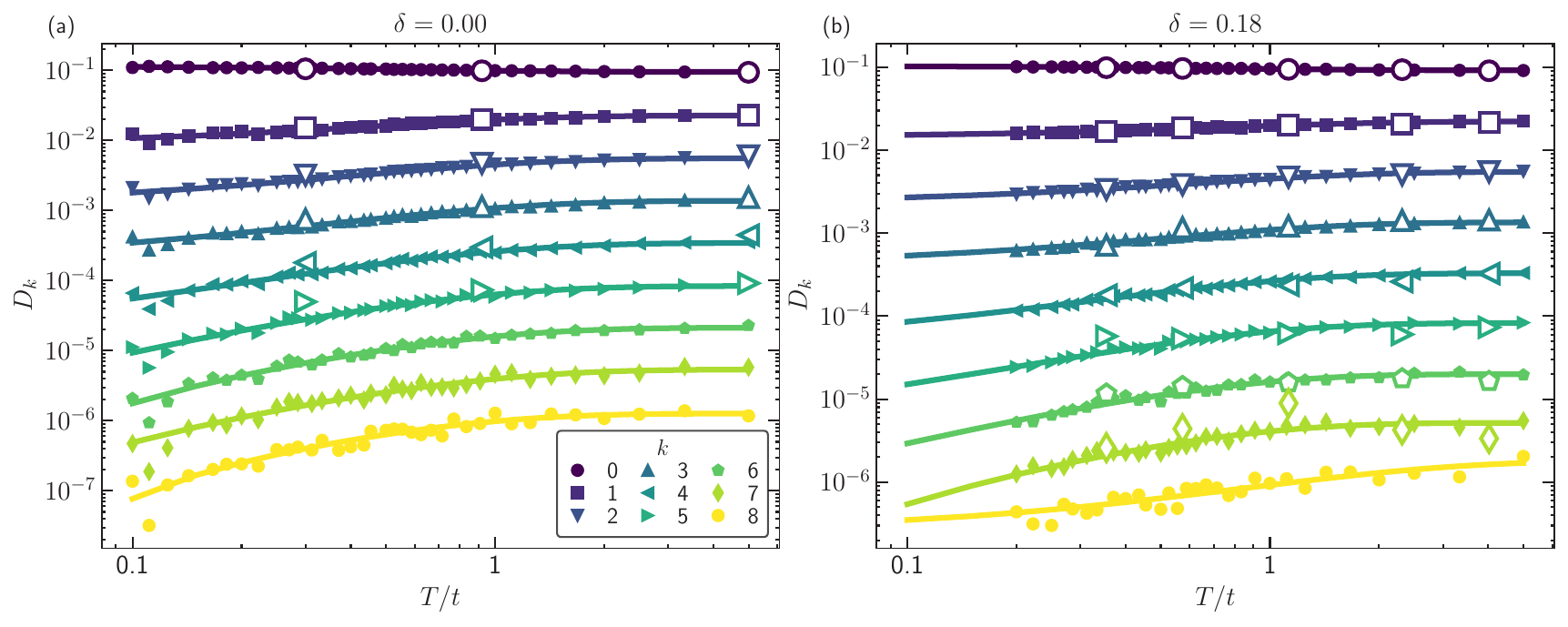}
\caption{\textbf{Dissimilarities ($D_k$) in the FHM at $U/t=8$ from spin-resolved snapshots.} (a)  $\delta=0$ (b)  $\delta = 0.18$. Solid markers correspond to simulated snapshots in $10\times10$ lattices and open markers to experimental data. Solid lines are exponential fits to the data. In both cases, $D_k$ exhibits an exponential decay with temperature (solid lines) and a rapid suppression with increasing $k$.}\label{fig::DkvsT} 
\end{figure*}

For both the complexities and the dissimilarities, results based on snapshots that are obtained experimentally using a quantum gas microscope~\cite{Khatami2020} (denoted by open markers) are in excellent agreement with the theory curves. Furthermore, this agreement between the numerical and experimental values of the dissimilarities holds for every $D_k$ at all temperatures considered, as evidenced in Fig.~\ref{fig::DkvsT} where we show $D_k$ vs $T/t$ for the half-filled case and the 18\% doped FHM. Such excellent agreement demonstrates the applicability of the method to current experiments involving ultracold atoms in optical lattices.

\begin{figure}[b]
\includegraphics[width=\linewidth]{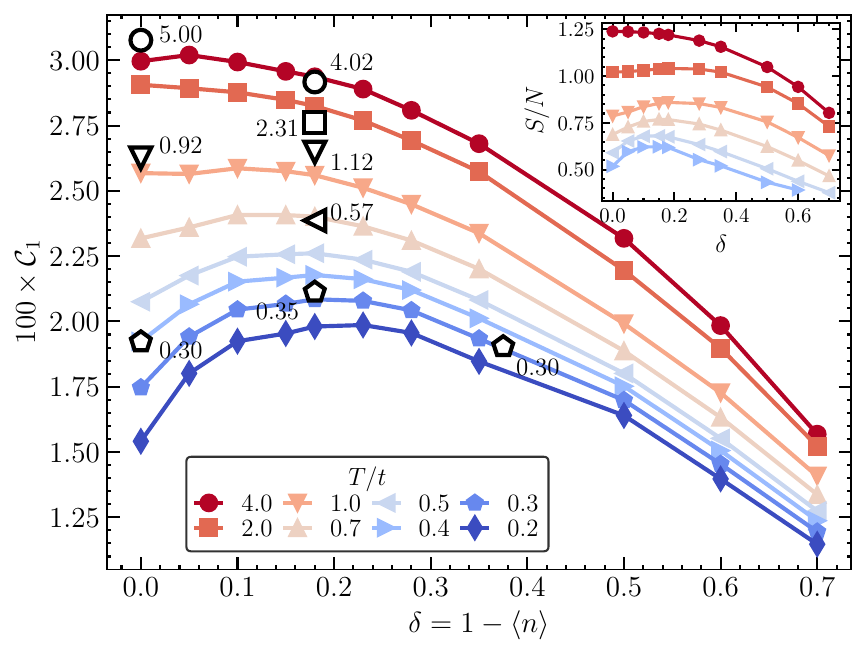}
\caption{\textbf{$\mathcal{C}_1$ vs $\delta$ for the FHM at $U/t=8$ from spin-resolved snapshots.} Solid markers correspond to simulated snapshots and open black markers to experimental data. Temperature estimates from the experiment are indicated next to each marker. Inset: Entropy per site as a function of doping from the numerical linked-cluster expansion (NLCE)~\cite{Khatami2011}.
}\label{fig::SCvsdoping} 
\end{figure}

More interestingly, we find that $\mathcal{C}_1$ vs temperature or doping exhibits the same trends as the entropy per site $S/N$ in the model (see Fig.~\ref{fig::SCvsdoping} for the doping dependence at different temperatures). 
We observe that below $T/t < 1$, the structural complexity develops a peak around 20\% doping. Although the peak in the entropy occurs at a slightly lower doping (see the inset of Fig.~\ref{fig::SCvsdoping}), the shape and trends are fairly similar. Additionally, we also observe good agreement with the experimental data at all temperatures considered, which further supports the validity of our results. Because of the correlation between $\mathcal{C}_1$ and $S/N$, this suggests the structural complexity of Fermi-Hubbard snapshots can be used as a quick and rapid proxy for the entropy of the system. This can be very useful in optical lattice experiments, where after loading the lattice, the calculation of the entropy is not straightforward~\cite{Cocchi2017}. 

It is worth mentioning that previous studies in one-dimensional systems: bit-strings~\cite{Sotnikov2022} and the extended Hubbard model~\cite{Xiao2022} have demonstrated that the structural complexity draws a close analog to the entanglement entropy since both capture the existence of nonlocal correlations in the models. In the present study we do not have access to the entanglement entropy and therefore only focus on the thermodynamic entropy. Although the entanglement entropy and the thermodynamic entropy are not equivalent to each other, they are associated with the development of correlations in the system and with the freezing of degrees of freedom, respectively, which generally occur together as we vary a tuning parameter. The establishment of connections between the structural complexity and the entanglement entropy in doped Fermi Hubbard models at finite-$T$ is an open question that requires further exploration.

Another important observation is the fact that at the temperatures considered, the region where the structural complexity is maximal corresponds to the strange metal region.
The broad peak in $\mathcal{C}_1$ spanning $\delta \in [10\%,30\%]$ is consistent with the doping value $\delta_c \sim 30\%$ that corresponds to the crossover to the Fermi liquid phase~\cite{Koepsell2021}. Moreover, the trends in the structural complexity vs doping illustrates that features related to phases that emerge in this doping range at lower temperatures (such as fluctuating stripes~\cite{Huang2018,Huang2023,Mai2023}) are more complex than the AFM parent state at half-filling, and that this metallic region is also more complex than the conventional Fermi liquid found at larger dopings. As optical lattice experiments evolve and stripe physics becomes available, we expect that signatures of the presence and periodicity of stripe orderings should be detectable by analyzing the contributions of the dissimilarities at different scales and $\mathcal{C}_1$.

\subsubsection{Local moments and total density}\label{sec::moments_densities}

\begin{figure}[tbp!]
\includegraphics[width=\linewidth]{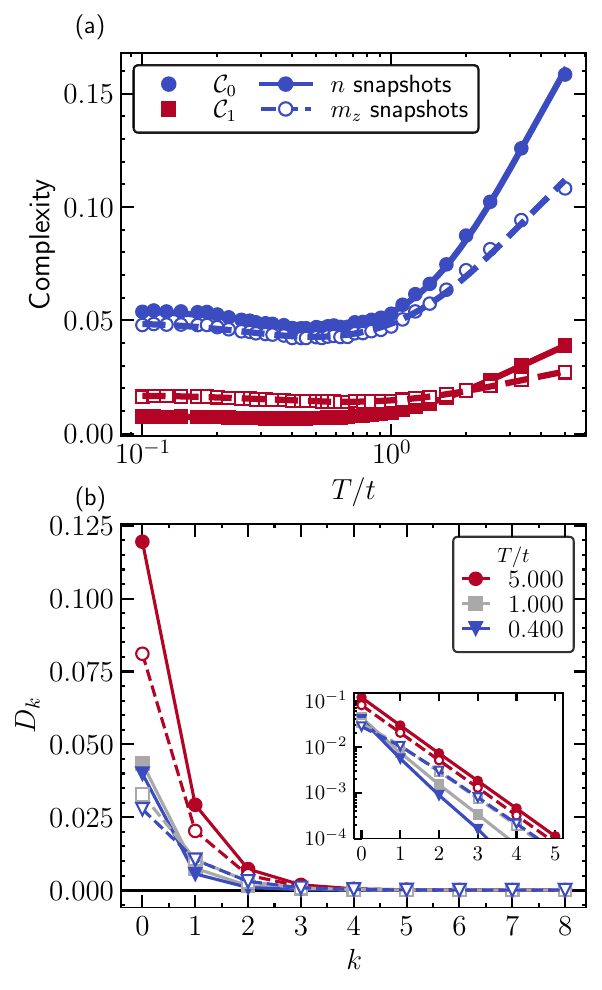}
\caption{\textbf{Structural complexity of the half-filled FHM at $U/t=8$ from simulated density and local moment snapshots.} 
(a) $\mathcal{C}_0$ (blue circles) and $\mathcal{C}_1$ (red squares) as a function of temperature for the local moments (open markers), and the total density (solid markers). Lines are guides to the eye. (b) Dissimilarities as a function of coarse-graining steps at different temperatures for local moments (open markers) and total density (solid markers). The inset presents the same data in log scale.}\label{fig::SC_m_n} 
\end{figure}

Before quantum gas microscopy had access to images of both fermionic species of the same system at the same time, density snapshots of the FHM could not be obtained; only local moments could be detected separately. Nowadays, with spin-resolved imaging, local moment as well as total density images can be created~\cite{Koepsell2020PRL,Hartke2020,Hartke2023}. In this section we study their structural complexities and dissimilarities.

In Fig.~\ref{fig::SC_m_n}(a) we show the structural complexities of the local moments (open markers and dashed lines) and the total density (solid markers and solid lines) of the half-filled FHM as a function of temperature. In contrast to what we observed for the spin-resolved snapshots, $\mathcal{C}_0$ is not constant but exhibits a non-monotonic behavior as a function of $T/t$. This is because unlike for single-species snapshots, as the temperature changes, the number of doublons and therefore the balance between filled and empty pixels in the snapshots changes too. Such non-monotonic behavior persists for $C_1$, albeit with smaller magnitude changes. 

We find that for $T/t \leq 2$, $\mathcal{C}_0$ for the density snapshots is larger than for the local moment snapshots, but the opposite is true for $\mathcal{C}_1$. 
We can understand this through the following argument: Below the characteristic AFM temperature, where the upturn in double occupancy starts as we lower $T$, for density snapshots, the doublon-hole fluctuations take place on nearest-neighbor sites, and so the first coarse-graining step already renders mostly uniform images. On the contrary, for local moment snapshots, because both holes and doublons show up as empty pixels, one still has significant structure for later coarse-graining steps. 

To complement these results, in Fig.~\ref{fig::SC_m_n}(b) we show the dissimilarities of the half-filled FHM as a function of coarse-graining steps at different temperatures. $D_k$ for both the local moments and total density exhibit an exponential decay with $k$ at fixed $T/t$. 

\begin{figure}[tbp!]
\includegraphics[width=\linewidth]{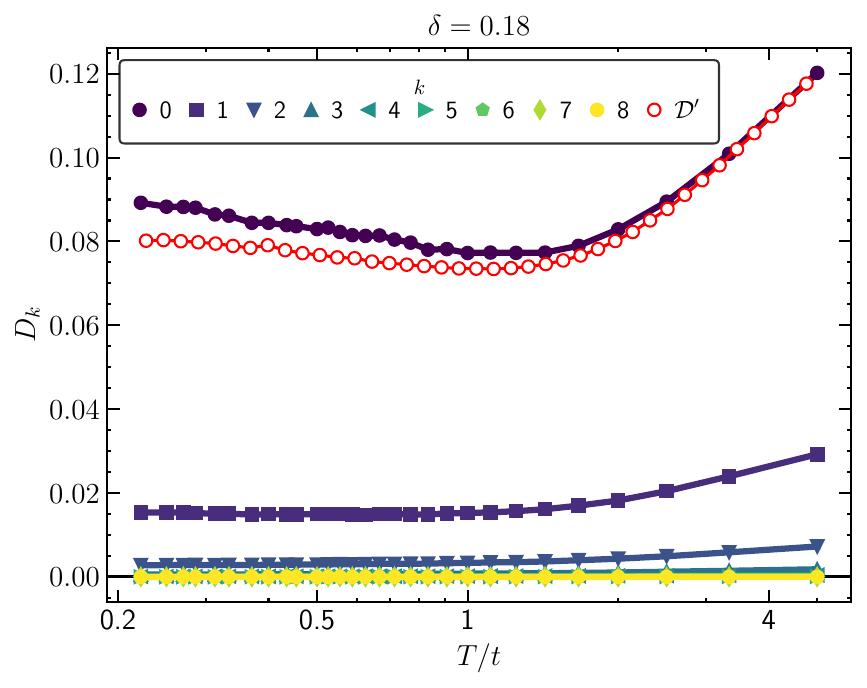}
\caption{\textbf{$D_k$ vs $T/t$ for the FHM with $U/t=8$ from simulated density snapshots at 18\% doping.} Red open markers come from the NLCE for $\mathcal{D}' = (3/4)[\mathcal{D} + 0.5\delta(1-\delta)]$.
}\label{fig::Dks_dens_doub} 
\end{figure}

\begin{figure*}[tbp!]
\includegraphics[width=\linewidth]{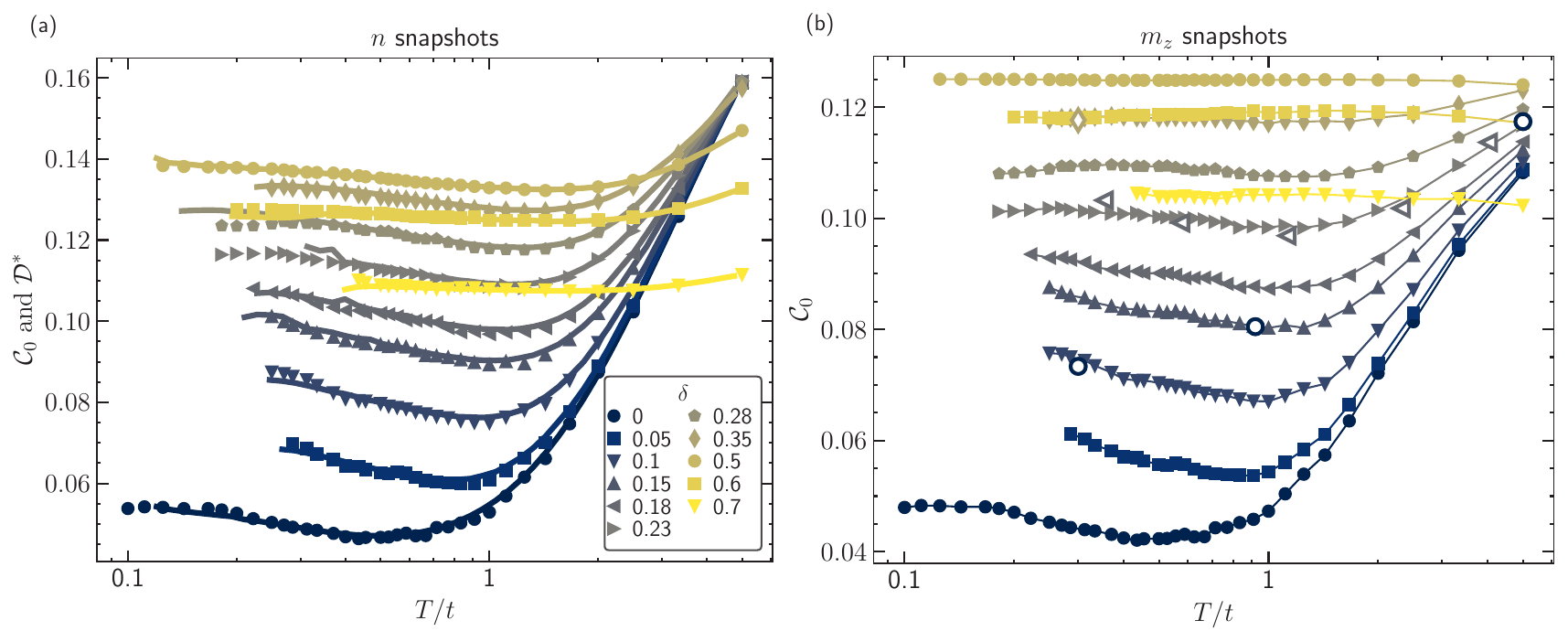}
\caption{\textbf{$\mathcal{C}_0$ vs $T/t$ for the FHM at $U/t=8$ from  density and local moment snapshots.} (a) Total density: Markers correspond to the structural complexity and lines to $ \mathcal{D}^* = \mathcal{D} + 0.5\delta(1-\delta)$ obtained with NLCE. (b) Local moments: Solid markers are based on simulated snapshots, open markers based on experimental ones, and lines are guides to the eye. }\label{fig::SC0_dens_doubs} 
\end{figure*}

In Fig.~\ref{fig::Dks_dens_doub}, we show the temperature dependence of the dissimilarities for the density snapshots at $\delta =0.18$. Here we observe that (i) $D_k$ decays rapidly with increasing $k$, (ii) $D_0$ exhibits a non monotonic behavior with $T/t$, akin to the double occupancy, $\mathcal{D} = \expect{n_{i\uparrow} n_{i\downarrow}}$, and (iii) it closely follows the function $\mathcal{D}' = (3/4)[\mathcal{D} + 0.5\delta(1-\delta)]$. This expression can be derived analytically under the assumption that the uncorrelated parts of the correlation functions dominate (see Appendix~\ref{App::D0}), and it holds for other dopings too. The agreement at high-$T$, where the assumption is valid, illustrates that $D_0$ for the density snapshots is in fact closely related to the double occupancy fraction in the model. More important are the discrepancies at $T/t \leq 1$, which signal the presence of non-trivial 2-point nearest- and next-nearest-neighbor spin-resolved density-density correlations in the model. A similar approximation for $D_0$ based on local moment snapshots proves challenging as 3-point and 4-point density correlators enter the analysis (see Appendix~\ref{App::D0}).

Fig.~\ref{fig::SC0_dens_doubs}(a) shows $\mathcal{C}_0$ as a function of $\delta$ for the density snapshots along with a shifted $\mathcal{D}$ from the numerical linked-cluster expansion (NLCE)~\cite{rigol2006numerical,tang2013short,Khatami2011} for each value of the doping. $\mathcal{C}_0$ exhibits an interesting behavior: (i) it is non-monotonic with temperature for all values of $\delta$, (ii) is largest at $\delta =0.5$, except at very high temperatures, and remarkably (iii) fully matches the shifted double occupancies at all temperatures and all dopings considered. Previously, we discussed that $D_0$ is closely correlated to the double occupancy fraction in the model. When higher order $D_k$'s are added, they capture the remaining features of the double occupancy, and therefore, $\mathcal{C}_0$ captures its temperature dependence extremely well. 

In contrast, the structural complexities of the local moment snapshots [Fig.~\ref{fig::SC0_dens_doubs}(b)] are more complicated to interpret. Although the behavior of the curves for all dopings except $\delta = 0.5$ look fairly similar to their counterparts based on density, we were unable to derive a simple analytic expression based on thermodynamic properties in this case due to the presence of higher order correlations even at the first coarse graining step. The most interesting region for this complexity occurs at $\delta =0.5$ where for $T/t \leq 1$, $\mathcal{C}_0 = 0.125$, the same constant value of the structural complexity as for the single-species snapshots at half-filling.
This region corresponds to the CDW phase observed at quarter-filling in the FHM~\cite{Cheuk2016,Khatami2020}, where one sublattice is occupied and other sublattice is empty, and therefore, $\mathcal{C}_0$ is merely capturing the classical complexity associated with having the largest entropy in the images with an equal number of empty and filled pixels.

Contrary to what occurred with the single-species snapshots, here we do not find excellent agreement between experiment and theory. Figure~\ref{fig::SC0_dens_doubs}(b) displays experimental data at different dopings as open markers. For $\delta = 0.35$ the open diamond indeed matches with the simulations at the same doping. However, $\mathcal{C}_0$ at $\delta =0.6$ has the same value.
On the other hand, for data at $\delta = 0.18$, the open left triangles match fairly well with the simulations at a slightly larger doping of $\delta = 0.23$, which might suggest the experimental data was taken at such higher doping. Finally, $\mathcal{C}_0$ based on experimental data at half-filling does not show good agreement with that calculated for the simulated images, and therefore could signal issues with the imaging procedure. Hence, the structural complexity of local moments can be used to further diagnose imaging issues.

\begin{figure*}[tbp!]
\includegraphics[width=\linewidth]{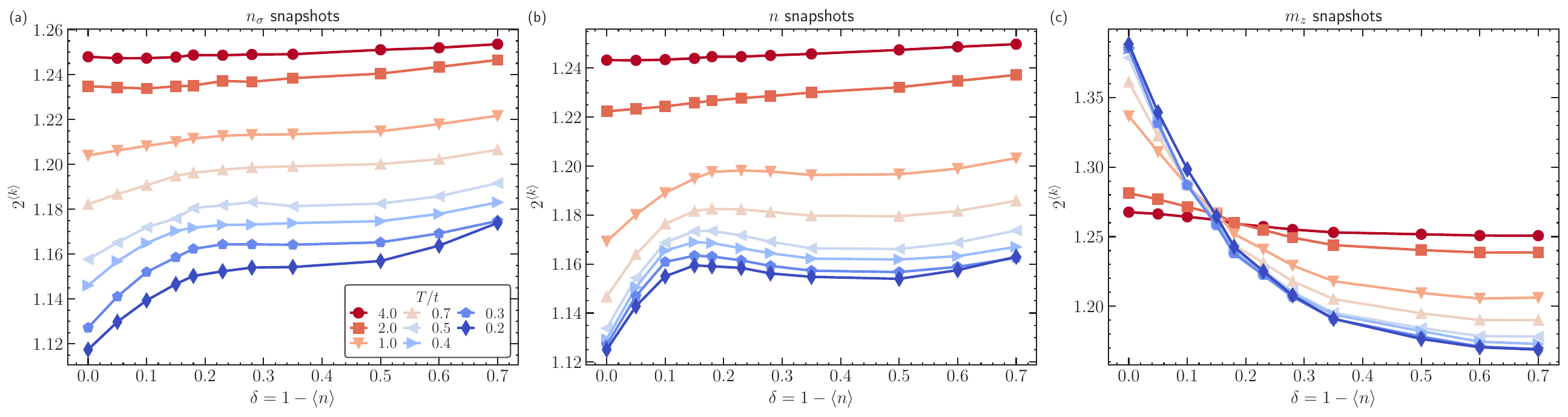}
\caption{\textbf{Average linear length scale obtained from dissimilarities vs doping.} (a) From single-species snapshots, (b) from total density snapshots, (c) from local moment snapshots.}\label{fig::kavg} 
\end{figure*}

Fig.~\ref{fig::kavg} presents $2^{\expect{k}}$ vs doping at different temperatures for the three different types of snapshots analyzed. At the highest temperature considered, $T/t=4$, $2^{\expect{k}}$ increases slowly with $\delta$ for the $n_\sigma$ and $n$ snapshots, but decreases for the $m_z$ snapshots. This trend persists down to all temperatures considered for the $m_z$ snapshots, becoming sharper as the temperature is lowered, but surprisingly exhibiting a crossing at $\delta \sim 18\%$, which points to a possible temperature-independent length scale in $m_z$ at this doping. This coincides with the region where structural complexity is maximal and corresponds to the strange metal region. Since this occurs at easily experimentally-accessible regimes, it is a phenomenon that can be explored immediately in the experiment. 

$2^{\expect{k}}$ vs temperature for $n_\sigma$ and $n$ snapshots also displays interesting behaviors at the same doping. As seen in Fig.~\ref{fig::kavg}(a) for $n_\sigma$, the curves develop a plateau that starts around $\delta \sim 18\%$ and extends up to $\delta \sim 40\%$, while for $n$ in Fig.~\ref{fig::kavg}(b) the curves develop a broad peak around $\delta \sim 18\%$. 

\section{Conclusions and Discussions}\label{sec::Conclusions}

In this paper, we show how the structural complexity, derived from overlaps of consecutive coarse-grainings of images can be used to analyze real space snapshots of strongly-correlated electronic systems to reveal their underlying physics. We benchmark our results for classical models of magnetism and show that the complexity measure can predict the magnetic moment and the critical temperature very well. We then turn to the FHM and compute the structural complexity for various types of snapshots generated via DQMC simulations or obtained through quantum gas microscopy in optical lattice experiments. Among other things, we find that when calculated for the total density snapshots, it matches the double occupancy as a function of temperature shifted by a density-dependent term. We identify specific contributions that correlate with the entropy of the system as a function of doping when dealing with snapshots of individual species of fermions. Dissimilarities of patterns at different coarse-graining steps also reveal relevant length scales in different regions and point to interesting behaviors near $18\%$, where the strange metal phase has been observed. Hence, the technique promises a ``theory-free'' method to extract phase changes or even physical properties not otherwise accessible based on experimental images alone. To that end, we have developed an open-source Python package~\cite{Ibarra2023_github}.

An immediate application of this technique can be for the study of Fermi-Hubbard models with larger spins and enhanced SU($N>2$) symmetries, which are expected to exhibit exotic behaviors with a complicated $N$-dependence~\cite{Corboz2011,Nataf2014,IbarraGarciaPadilla2021,IbarraGarciaPadilla2023,Feng2023}. In recent years, experiments with alkaline-earth-like-atoms in optical lattices have studied the SU($N$) FHM~\cite{Taie2012,Hofrichter2016,Ozawa2018,Taie2020,tusi2022flavour,Pasqualetti2024}, and future implementations of quantum gas microscopes for two-dimensional optical lattices will provide spin-resolved projective measurements of the SU($N$) FHM, which can be analyzed using the multiscale structural complexity to extract valuable physics. Furthermore, we expect the technique to also be useful to analyze other types of ultracold atom experiments images, such as absorption imaging of interferometric patterns~\cite{Kasia1,Kasia2}, and self-simmilarities of atomic wavepackets~\cite{Artoni2009}.

In addition, future work will explore the possibility of extending the structural complexity measure to other lattice geometries. This is motivated by the field's interest in studying frustration in triangular~\cite{Xu2022,Jirayu2022,Lebrat2023,prichard2023} and kagome~\cite{Jo2012} lattices with ultracold atoms, as well as the exotic physics that arises due to the non-periodicity in quasicrystals~\cite{Viebahn2019}, and the flexibility of engineering arbitrary lattice geometries using optical tweezer arrays~\cite{Yan2022,Haotian2023}.

A particularly intriguing application would be to employ the structural complexity to study time-independent and time-dependent Green’s functions,
since the dynamical properties of models used to describe strongly-interacting matter such as the Hubbard, Holstein, and Su–Schrieffer–Heeger Hamiltonians
are much more challenging to explore using conventional numerical approaches.

\begin{acknowledgments}
We are grateful to Waseem Bakr, Elmer Guardado-Sanchez, and Benjamin M. Spar for sharing their experimental snapshots for the Fermi-Hubbard model with us. We also thank Waseem Bakr, Martin Zwierlein for their useful comments on our manuscript.
The authors are supported by the grant DE-SC-0022311, funded by the U.S. Department of Energy, Office of Science. EIGP would like to acknowledge useful discussions with Steven Johnston and Bejamin Cohen-Stead while visiting the University of Tennessee, Knoxville.
\end{acknowledgments}

\appendix

\section{Expression for $\mathcal{C}_0$ in the infinite temperature limit}\label{App::High_T}

In this appendix, we derive the structural complexity for the Ising models in the fully disordered phase, i.e. where each spin orientation is equally probable, given by Eq.~\eqref{eq::C0_highT}. 

Let us calculate $D_0$ first. For the $2\times 2$ coarse-graining window, we have $2^4 =16$ equally probable combinations of $1$'s and $-1$'s. After the first coarse-graining step, the coarse-grained values obey a binomial distribution. These are presented in Table~\ref{table::D0_T}, as well as the values and probabilities of the overlaps $\vert O_{1,0} - \frac{1}{2} \left( O_{1,1} + O_{0,0}\right)\vert$.
\begin{table}[h!]
\begin{center}
\begin{tabular}{ |c|c|c|c| } 
\hline
\multicolumn{2}{|c|}{\textbf{Coarse-grained}} & \multicolumn{2}{|c|}{\textbf{Overlaps}} \\
\hline
Value & Probability & Value & Probability \\
\hline 
$\pm$ 1 & 1/16 & 0 & 1/8\\
$\pm$ 1/2 & 1/4 & 3/8 & 1/2 \\
0 & 1/8 & 1/2 & 3/8 \\
 \hline
\end{tabular}
\caption{Possible values for the first coarse-graining step and associated probabilities, as well as the values and probabilities of the overlaps for the fully disordered spin snapshots.}\label{table::D0_T}
\end{center}
\end{table}

With these results we can calculate the average $D_0 = \sum_n p_n d_n$, where $d_n$ corresponds to the value of the overlap, and $p_n$ its probability. In this case,
\begin{equation}
    D_0 = \left(0 \times \frac{1}{8}\right) + \left(\frac{1}{2} \times \frac{3}{8}\right) + \left(\frac{3}{8} \times \frac{1}{2}\right) = \frac{3}{8}.
\end{equation}

In order to calculate $D_1$, we now have to consider $5^4=625$ possible combinations for each window. After the second coarse-graining step, the values and overlaps obey the probabilities presented in Table~\ref{table::D1_T}.

\begin{table}[h!]
\begin{center}
\begin{tabular}{ |c|c|c|c| } 
\hline
\multicolumn{2}{|c|}{\textbf{Coarse-grained}} & \multicolumn{2}{|c|}{\textbf{Overlaps}} \\
\hline
Value & Probability & Value & Probability \\
\hline 
$\pm$ 1 & 1/65536 & 0 & 905/32768\\
$\pm$ 7/8 & 1/4096 & 3/128 & 329/2048 \\
$\pm$ 3/4 & 15/8192 & 1/32 & 111/1024 \\
$\pm$ 5/8 & 35/4096 & 1/16 & 9/64 \\
$\pm$ 1/2 & 455/16384 & 11/128 & 207/1024 \\
$\pm$ 3/8 & 273/4096 & 3/32 & 239/4096 \\
$\pm$ 1/4 & 1001/8192 & 1/8 & 123/4096 \\
$\pm$ 1/8 & 715/4096 & 19/128 & 39/512 \\
 0 & 6435/32768 & 5/32 & 9/128 \\
& & 3/16 & 21/512 \\
& & 27/128 & 83/2048 \\
& & 1/4 & 27/4096 \\
& & 35/128 &  9/512 \\
& & 9/32 &  9/1024 \\
& & 5/16 &  3/512 \\
& & 43/128 &  3/2048 \\
& & 11/32 &  9/4096 \\ 
& & 3/8 & 1/8192 \\ 
& & 51/128 & 3/2048 \\
& & 1/2 & 3/32768 \\
 \hline
\end{tabular}
\caption{Possible values for the second coarse-graining step and associated probabilities, as well as the values and probabilities of the overlaps for the fully disordered spin snapshots.}\label{table::D1_T}
\end{center}
\end{table}

With these results we calculate the average $D_1$ as
\begin{equation}
    D_1 = \frac{3}{32} = D_0 \times \left(\frac{1}{4}\right).
\end{equation}
It is then easy to prove that the next dissimilarities involved in computing the structural complexity are given by 
\begin{equation}
    D_k = D_0 \times \left(\frac{1}{4^k}\right),
\end{equation}
and therefore we arrive at Eq.~\eqref{eq::C0_highT}, 
\begin{equation}
    \mathcal{C}_0 = \frac{3}{8} \left(\sum_{k=0}^\infty \frac{1}{4^k} \right) = \frac{3}{8} \left(\frac{4}{3}\right) = \frac{1}{2}.
\end{equation}  

\section{Expressions for $D_0$ for the density and local moment snapshots}\label{App::D0}

To simplify reading the following equations, in this appendix we use two indices $ij$ to label a site in an image. 

Now let us consider a two-dimensional image with linear dimension $L$ and $N= L\times L$ sites. On each site, given by the coordinate pair $ij$, the value of the image is $u_{ij}$. The first term $D_0$, is given by
\begin{align}
D_0 = \frac{1}{4N}  \bigg\vert  \sum_{i,j=1}^{L/2} &\bigg[ u_{2i-1,2j-1}(u_{2i-1,2j} + u_{2i,2j-1}) \nonumber \\
&+ u_{2i,2j-1}(u_{2i-1,2j} \nonumber \\
&+ u_{2i,2j}) + u_{2i,2j}(u_{2i-1,2j-1} + u_{2i-1,2j}) \bigg] \nonumber \\
& - \frac{3}{2} \sum_{i,j=1}^L u_{ij}^2 \bigg\vert
\end{align}
and it captures all correlations within a unit window, i.e. on-site, nearest and next-nearest neighbors.

In the case of density snapshots the images correspond to $u_{ij} = n_{ij,\uparrow} + n_{ij,\downarrow}$. So, $D_0$ is given by,
\begin{align}
    D_0 &= \frac{1}{4N}  \bigg\vert  2\bigg[\expect{n_{\uparrow}n_{\uparrow}}_{nn} +  \expect{n_{\uparrow}n_{\downarrow}}_{nn}\bigg] + \\ 
    &\qquad \quad \bigg[ \expect{n_{\uparrow}n_{\uparrow}}_{nnn}+  \expect{n_{\uparrow}n_{\downarrow}}_{nnn} \bigg] 
   - \frac{3}{2} \rho - 3 \mathcal{D} \bigg\vert\nonumber,
\end{align}
where we exploited the translational and rotational symmetry, and $\expect{n_{\sigma}n_{\tau}}_{nn}$ is shorthand for nearest-neighbor density-density correlations for spin $\sigma$ and $\tau$. The subindex $nnn$ indicates the next-nearest-neighbor correlator. When the uncorrelated part of these correlation functions dominates (for example at high-$T$), then $\expect{n_{\uparrow}n_{\uparrow}}_{nn} = \expect{n_{\uparrow}n_{\downarrow}}_{nn} = \expect{n_{\uparrow}n_{\uparrow}}_{nnn} =\expect{n_{\uparrow}n_{\downarrow}}_{nnn} = \rho^2/4$, and so,
\begin{equation}
    D_0 \approx \frac{1}{4} \bigg\vert \frac{3}{2} \rho^2 - \frac{3}{2} \rho - 3 \mathcal{D} \bigg\vert =  \frac{3}{4} \bigg\vert \frac{1}{2} \delta (1-\delta) - \mathcal{D} \bigg\vert.
\end{equation}
This final expression indicates that the double occupancy, shifted by an amount $0.5 \delta(1-\delta)$ and rescaled, should agree with $D_0$.

In the case of the local moments, the images correspond to $u_{ij} = (n_{ij,\uparrow} - n_{ij,\downarrow})^2$. Working out the expressions, and under the same assumptions of translational, rotational, spin permutation symmetry, and that the correlated part of the functions dominate, we find,
\begin{equation}
    D_0 \approx \frac{1}{4} \bigg\vert \frac{3}{2} \frac{1}{4}\rho^2(2-\rho^2) - \frac{3}{2} m_z \bigg\vert =  \frac{3}{8} \bigg\vert \frac{1}{4} (1-\delta^2)^2 - m_z \bigg\vert.
\end{equation}

We expect that the disagreement between this $D_0$ and any simple function of $m_z$ is related to the fact that 3-point and 4-point density correlators enter the analysis, i.e., the assumption that the uncorrelated parts dominate is no longer valid.

\bibliography{SC_FHM.bib}

\end{document}